\documentclass{IEEEtran}
\usepackage{cite}
\usepackage{amsmath,amssymb,amsfonts}
\usepackage{algorithmic}
\usepackage{graphicx}
\usepackage{textcomp}

\usepackage{amsmath,amsfonts}

\usepackage[linesnumbered,ruled]{algorithm2e}

\usepackage{array}
\usepackage[caption=false,font=normalsize,labelfont=sf,textfont=sf]{subfig}
\usepackage{textcomp}
\usepackage{stfloats}
\usepackage{url}
\usepackage{verbatim}
\usepackage{graphicx}
\usepackage{cite}
\usepackage{multirow}
\usepackage{amssymb} 
\usepackage{cases} 
\usepackage{makecell} 
\usepackage{diagbox} 
\usepackage{hyperref}
\usepackage{xcolor}
\usepackage{svg}

\usepackage{booktabs}   
\usepackage{tabularx}   

\newcolumntype{Y}{>{\centering\arraybackslash}X}

\def\BibTeX{{\rm B\kern-.05em{\sc i\kern-.025em b}\kern-.08em
		T\kern-.1667em\lower.7ex\hbox{E}\kern-.125emX}}
	
\begin{document}
\title{{Emulation of Closely Spaced Spatial Sensing Targets Using a Parabolic Cylindrical Compact Range with a Multi-Feed Amplitude-Phase Controlled Linear Array}}
\author{Zhangzhang Jiang, Guokai Jiang, Le Yu, Chunhui Li, Zhengpeng Wang, and Wei Fan
	\thanks{Zhangzhang Jiang, Le Yu and Chunhui Li are with the National Mobile Communications Research Laboratory, School of Information Science and Engineering, Southeast University, Nanjing 210096, China (e-mail:\{jiangzhangzhang,yule,lichunhui\}@seu.edu.cn)
	
	Guokai Jiang is with the China Automotive Technology and Research Center Co., Ltd, Tianjin 300300, China (e-mail: jiangguokai@catarc.ac.cn).
	
	Zhengpeng Wang is with the Electronics and Information Engineering, Beihang University, Beijing 100191, China (e-mail: wangzp@buaa.edu.cn).
	
	Wei Fan is with the National Mobile Communications Research Laboratory, School of Information Science and Engineering, Southeast University, China, and also with Purple Mountain Laboratories, China  (e-mail:\{weifan\}@seu.edu.cn). \textit{Corresponding author: Wei Fan.} }
	
}
	
\maketitle
	
\begin{abstract}
	Integrated sensing and communication (ISAC) base stations (BSs) are emerging as critical infrastructure for 6G networks, requiring advanced over-the-air (OTA) testing methodologies that can emulate dynamic, multi-target scenarios under far field (FF) conditions. However, existing testing solutions often suffer from limited angular flexibility, insufficient bandwidth and prohibitive costs. {To address these challenges, this paper proposes an OTA testing system that integrates a radar target simulator (RTS), a multi-port amplitude and phase modulation (APM) network, and a multi-feed linear array with a single parabolic cylindrical compact range (SPCCR).} The proposed system is capable of emulating closely-spaced spatial targets with varying incident azimuth angles. This setup meets the essential conditions for ISAC BS testing, including FF characteristics, wideband and multi-band compatibility, and the capability for sensing multiple dynamic closed-spaced targets. The feasibility of the proposed method is validated through both FEKO simulation and measurement. Experimental results demonstrate that the system successfully emulates multiple closely spaced targets within an azimuth range of $\pm 10^\circ$ with high emulation accuracy, while maintaining excellent quiet zone (QZ) performance.
\end{abstract}


\section{Introduction}
\label{sec:introduction}

\IEEEPARstart{W}{ith} the rapid advancement of sixth-generation (6G) technology, integrated sensing and communication (ISAC) base stations (BSs) have emerged as the core infrastructure for next-generation wireless networks \cite{wang2023road}. The ISAC BSs are expected to simultaneously enable high-rate data communication and high-precision environmental sensing capabilities comparable to radar systems, utilizing advanced phased array architectures to support beam scanning and target detection \cite{paul2016survey,li2023wide,topal2024multi,11169758}. While current ISAC standardization primarily focuses on specific frequency bands, notably 4.9 GHz and 26 GHz, the broader radar community routinely exploits a comprehensive spectrum, spanning S, X, Ku, and Ka bands \cite{feng2020joint}. Diverse frequency bands and expanded system bandwidths might be explored for future ISAC technique. ISAC BSs typically adopt multiport, fully connected array structures with tightly integrated RF front ends and antenna modules \cite{cao2024sensing}. Moreover, the design of ISAC BSs is moving from analog beamformer to digital beamformer phased array structure with increased array aperture, which will significantly improve the spatial resolution of the system \cite{10158711}.

Consequently, the evolution of these advanced architectures in ISAC BSs requires corresponding advanced testing methods, which introduces new challenges to current BS testing methodologies, rendering conventional testing system inadequate for comprehensive performance validation of ISAC BSs. Therefore, establishing an effective test system for ISAC BSs becomes imperative, which entails addressing a set of core requirements:
\begin{itemize} 
	\item 
	First, the test system must provide a far field (FF) environment within a quite zone (QZ) enclosing the ISAC BS of large antenna apertures with plane wave illumination.
	
	\item 
	Second, the test system must support wideband frequency coverage, potentially covering frequency bands up to mmWave bands, alongside sufficient bandwidth capacity, due to the requirement of high-rate data communication and fine sensing resolution of ISAC BSs. 
	
	\item 
	Third, the test system must be capable of emulating dynamic multiple radar targets with fully configurable attributes, including radar cross section (RCS), range, Doppler, and angle of arrival (AoA). 
	
	\item 
	Fourth, the test system must have the ability to generate closely-spaced spatial targets to rigorously evaluate the ISAC BS's angular resolution capability (e.g. $\le 3.5^{\circ}$). 
	
	\item 
	{Finally, achieving all these capabilities within a compact infrastructure for practical deployment.}
	
\end{itemize}
It is worth noting that the fourth requirement is very challenging but highly important for ISAC BS testing. {Since the spatial resolution of ISAC BS is significantly enhanced by the large antenna aperture and the transition to digital beamforming architectures, the ISAC BS is required to have certain spatial resolution for sensing applications \cite{liu2022survey}. However, due to the RF chain imbalances and inhomogeneous element radiation patterns in real world deployments, the practical spatial resolution often degrades \cite{10474102} \cite{lu2024integrated}.} Therefore, we have to emulate closely-spaced spatial targets scenario for ISAC BS testing to ensure the designed spatial discrimination capabilities of ISAC BS can be achieved in practical designs.

In our previous work, Li et al. \cite{11306233} proposed a conducted testing scheme for Sub-6 GHz ISAC BS evaluation, which relies on the amplitude and phase modulation (APM) network and the radar target simulator (RTS) to emulate multiple targets. However, the highly integrated design in modern ISAC BSs often precludes access to test ports \cite{liu2024senscap}, rendering conducted schemes impractical.
\begin{table*}[htbp]
	\centering
	\caption{Comparison of Existing Target Emulation Methodologies}
	\label{tab:comparison_academic}
	
	\small 
	\setlength{\tabcolsep}{3pt} 
	\renewcommand{\arraystretch}{1.3} 
	
	\begin{tabularx}{\textwidth}{
			>{\centering\arraybackslash}c  
			>{\centering\arraybackslash}c  
			>{\centering\arraybackslash}X  
			>{\centering\arraybackslash}c  
			>{\centering\arraybackslash}c  
			>{\centering\arraybackslash}c  
			>{\centering\arraybackslash}X  
		}
		\toprule
		\textbf{Reference} & \textbf{Category} & \textbf{Methodology} & \textbf{\makecell{Angular Target \\ Emulation}} & \textbf{\makecell{Angular Target \\ Resolution}} & \textbf{\makecell{Test Freq. \\ Band}} & \textbf{Limitations} \\ 
		\midrule
		
		\cite{11306233} & Conducted & APM + RTS & Multiple angles & High & Sub-6 GHz & No access to test ports \\
		
		\cite{Noffz_CATR} & OTA & CATR + RTS & Single angle & Low & No limitations & Lack spatial emulation \\
		
		\makecell{\cite{buddappagari2021over}, \cite{asghar2021radar}} & OTA & \makecell{Mechanical probe \\ movement} & Multiple angles & Medium & 77, 79 GHz & Low angular flexibility \\
		
		\makecell{\cite{scheiblhofer2017low}, \cite{gadringer2018radar}}  & OTA & \makecell{Electronic \\ switching} & Multiple angles & Medium & 24, 77 GHz & Massive number of probes \\
		
		\cite{diewald2021arbitrary} & OTA & Field synthesis & Multiple angles & High & 77 GHz & FF assumption + narrowband \\
		
		\cite{maaskant2021new} & OTA & Waveguide PWG & Multiple angles & Limited & 5 GHz & Massive number of probes \\
		
		\makecell{\cite{Keysight_AD1012A}, \cite{NMT}} & OTA & \makecell{Massive miniature \\ RTS array} & Multiple angles & High & Customization & Extremely high cost \\
		\midrule
		
		\textbf{Our work} & \textbf{OTA} & \textbf{\makecell{SPCCR + Linear \\ array + APM + RTS}} & \textbf{Multiple angles} & \textbf{High} & \textbf{{2 $\sim$ 4 GHz}} & \textbf{{Only in azimuth plane}} \\
		\bottomrule
	\end{tabularx}
\end{table*}
Nevertheless, current mainstream OTA testing schemes also suffer from fundamental limitations, which are unable to meet the testing requirements of ISAC BSs. The compact antenna test range (CATR) radar test system proposed in \cite{Noffz_CATR} combines the CATR with RTS, which can only generate radar target from one single direction. The virtual road simulation and test area (VISTA) provides the capability to emulate targets from different angles \cite{buddappagari2021over} \cite{asghar2021radar}, but it relies on mechanical movement of probe antennas to achieve specific AoA, which cannot realize dynamic angular domain adjustment and it might not support ISAC BS with large aperture due to violation of FF assumption. Scheiblhofer et al. \cite{scheiblhofer2017low} and Gadringer et al. \cite{gadringer2018radar} employed electronically switched front-ends to achieve multiple radar targets simulation. However, these approaches require a high density of front-ends to maintain sufficient angular resolution. To address the resolution limitation, Diewald et al. \cite{diewald2021arbitrary} proposed a signal superposition approach to synthesize arbitrary angles without increasing the number of front-ends. However, this method is essentially narrowband and FF assumption should be met for the probe antenna. Maaskant et al. \cite{maaskant2021new} introduced a plane wave generator (PWG) placed inside an overmoded waveguide, which can synthesize oblique incident plane waves for device under test (DUT) in the near field (NF). Nevertheless, the requirement for half wavelength probe spacing in PWG will result in a massive number of antenna elements at higher frequencies required for ISAC, leading to high system cost. Besides, both the achievable angular resolution and the effective test-zone size are fundamentally constrained by the physical aperture of the waveguide. State-of-the-art commercial solutions, such as the Keysight Radar Scene Emulator \cite{Keysight_AD1012A} and NSI-MI’s real-time RF scene generators \cite{NMT}, employ a massive array of miniature probe antennas to emulate complex scenarios with multiple objects from different angles. These systems typically utilize RF switching matrices to route a limited number of RTS backend channels to selected active probes, subsequently employing highly accurate wideband phase shifters to synthesize target locations via the three-element method \cite{kojic1998modelling}. However, due to the compact test range, the FF assumption for the probe antennas is often violated. For different test frequency bands, different sets of probe antennas will be used. Consequently, the cost of these systems are extremely high.

\begin{table*}[htbp]
	\centering
	\caption{Comparison of CATR system in Literature}
	\label{tab:reflector_comparison}
	\small 
	\renewcommand{\arraystretch}{1.4} 
	
	\renewcommand{\tabularxcolumn}[1]{m{#1}} 
	
	\newcolumntype{Y}{>{\centering\arraybackslash}X}
	
	\begin{tabularx}{\textwidth}{c >{\hsize=1.1\hsize}Y >{\hsize=1.1\hsize}Y >{\hsize=0.8\hsize}Y >{\hsize=1.0\hsize}Y}
		\toprule
		\textbf{Reference} & \textbf{Type of reflector} & \textbf{Type of feed} & \textbf{Frequency range} & \textbf{Functionality} \\
		\midrule
		
		\makecell{\cite{johnson1969compact}} & Paraboloidal reflector & open-ended waveguide & 8.2 $\sim$ 12.4 GHz & Antenna testing, RCS testing \\
		\cmidrule(lr){1-5}
		
		\makecell{\cite{parini2016optimizing}} & Sector-shaped single offset reflector & 3$\times$3 waveguide horn & X-band & Enlarge the QZ area \\
		\cmidrule(lr){1-5}
		
		\makecell{\cite{9360112}} & Parabolic reflector & 3$\times$3 double slot corrugated horn & mmWave & Form multiple QZ with different deflection angles \\
		\cmidrule(lr){1-5}
		
		\makecell{\cite{yang2024enhancing}} & Single parabolic cylindrical reflector & 44 vivaldi antenna linear array & 2 $\sim$ 4 GHz & Enhance aperture efficiency for antenna testing \\
		\cmidrule(lr){1-5}
		
		\makecell{\cite{boswell1978parabolic}} & Tours reflector & 11 horn antenna & 10 GHz & Satellite communication \\
		\cmidrule(lr){1-5}
		
		\makecell{\cite{rowell2020multiple}} & 4 Blended rolled edge paraboloid reflectors & Dual-polarized antenna & FR2 & Radio resource management \\
		\cmidrule(lr){1-5}
		
		\makecell{\cite{kildal2018measurements}, \cite{razavi2018characterization}} & Cylindrical reflector & Linear array of dual-polarized antennas & 1.6 $\sim$ 2.7 GHz & Full-vehicle antenna testing \\
		\cmidrule(lr){1-5}
		
		\makecell{\cite{tafertshofer2024radar}} & Parabolic reflector & 2 shorted waveguide Tx antenna, 16 Rx & 76 $\sim$ 78 GHz & Radar target generation \\
		\midrule
		
		\textbf{Our work} & \textbf{Single parabolic cylindrical reflector} & \textbf{{44 vivaldi antenna linear array}} &  \textbf{{2 $\sim$ 4 GHz}} & \textbf{ISAC BS evaluation} \\
		\bottomrule
	\end{tabularx}
\end{table*}

As an indirect FF testing method, the CATR is specified in the 3GPP standards \cite{3gpp.38.810.v16.7.0} that can create the FF environment using a transformation with a parabolic reflector. Since the initial proposal of the CATR with a single feed and a paraboloidal reflector \cite{johnson1969compact}, numerous studies have been carried out in recent years \cite{zhang2024multiprobe}. Parini et al. \cite{parini2016optimizing} employed a 3 $\times$ 3 element array feed to improve and enlarge the QZ area. Ning et al. \cite{9360112} presented a multi-feeds compact range system, which can generate plane waves with nine incident angles (with a $3^{\circ}$ interval) at the QZ in millimeter wave frequency band. Yang et al. \cite{yang2024enhancing} utilized the linear array feed to enhance the aperture efficiency of single parabolic cylindrical compact range (SPCCR). In addition to advancements in feed architectures, several studies have also focused on improving reflector design. In \cite{boswell1978parabolic}, a parabolic torus reflector was proposed where multiple horn antennas illuminate the reflector, and each horn is rotated by a defined angle to synthesize multiple beams for satellite communication. The multiple CATR reflector system proposed by Rohde \& Schwarz \cite{rowell2020multiple} can generate at most four planar wavefronts with different incidences by deploying four independent CATR reflectors with respective feed antennas arranged on a planar semi-circle arc. The random line-of-sight (RanLOS) CATR measurement system consist of a cylindrical reflector fed by a linear array of dual-polarized antennas, which can generate a single plane wave toward the DUT \cite{kildal2018measurements}, \cite{razavi2018characterization}, while the DUT is mounted on a rotary positioner, allowing it to be rotated to different orientations. Moreover, Taftershofer et al. \cite{tafertshofer2024radar} presented a ray-tracing simulation approach for an automotive compact radar test range, that can generate two targets separated in azimuthal direction to test the angular resolution of a radar under test.

Despite these advancements in CATR architectures, existing solutions remain insufficient for the emerging requirements of ISAC BS testing. Most conventional methods rely on mechanical rotation of the DUT or bulky multi-reflector setups to change the AoA, which lacks the flexibility to generate simultaneous, closely spaced incident plane waves. This limitation significantly hinders the comprehensive performance evaluation of ISAC BSs. Yet, to date, no existing testing solution is capable of realizing high-precision emulation and verification of this capability under FF conditions. This technological gap renders the closed-loop evaluation of sensing performance in complex scenarios unattainable. Consequently, there is an urgent demand for an OTA testing paradigm that satisfies multi-target emulation, and validation capabilities for high angular resolution. To bridge this gap, we propose a closely spaced sensing target emulation algorithm for ISAC BSs testing using SPCCR with linear array feed. The comparative summary of the proposed framework with existing target emulation methods and the application of CATR system are shown in Table \ref{tab:comparison_academic} and \ref{tab:reflector_comparison}, respectively.

The main contributions of this paper are listed as follows:
\begin{itemize} 
	\item Firstly, we propose an innovative OTA test system specifically designed to support multi-target scenarios for ISAC BSs. {We introduce the RTS and APM network with multiple input ports into the SPCCR with multi-feed linear array. This system enables the emulation of closely-spaced spatial targets with different incident azimuth angles under FF conditions in compact setups}, supporting various frequency bands, wide system bandwidth, and extremely high accuracy for the QZ. While all PWG techniques are fundamentally constrained in supporting wideband frequency range with large instenous frequency bands \cite{11162628}.
	
	\item Additionally, the high surface precision of the SPCCR reflector supports an ultra-wide operating frequency range from $1$ GHz to $110$ GHz, while the current APM network and RTS support up to 40 GHz. Higher frequency bands testing can be achieved by updating the corresponding APM network, RTS and feed array, without changing the reflector. Compared to the conventional CATR with mechanical rotation, our proposed test system can realize electronically controlled AoA synthesis for multiple spatial targets.

	\item Finally, we validate the proposed framework through both full wave simulation and experimental validations. Simulations are conducted in FEKO to comprehensively illustrate the complete target emulation process. Subsequently, the methodology is validated using measured transfer functions to confirm its practical viability. The results both show that the system can accurately emulate closely spaced targets within $\pm 10^\circ$ degrees in the azimuth domain while maintaining excellent QZ performance, proving its suitability for ISAC BS testing.

\end{itemize}

The remainder of this article is organized as follows. Section II introduces the specific framework of our proposed OTA test systems and the optimization method for multiple targets emulation will be introduced in Section III. The FEKO simulation and experimental validations are provided in Section IV and V, respectively. All findings are finally concluded in Section VI.

\textit{Notations}: Bold uppercase characters $\mathbf{X} $ denote matrices; bold lowercase characters $\boldsymbol{x}$ denote vectors; $\mathbb{C}^{M \times N}$ represents the space of $M \times N$ complex-valued matrices. $j = \sqrt{-1}$ denotes the imaginary unit; $t$ and $f$ denote the time and frequency, respectively. $\|\cdot\|_2$ denotes the Euclidean norm ($L_2$-norm) of a vector; The symbol $\odot$ denotes the Hadamard (element-wise) product.


\section{Framework}

\subsection{Problem Statement}
Our objective is to emulate closely spaced multiple sensing targets for the ISAC BS under test, characterized by distinct RCS, range, system frequency band, Doppler, and AoA parameters. This capability enables comprehensive performance evaluation and optimization of ISAC BSs under dynamic multi-target scenarios. While existing RTS have achieved maturity in generating precise RCS, range, and Doppler signatures \cite{Keysight_E8718A} \cite{dSPACE_DARTS9040GT}, accurately emulating multiple targets in the spatial (angular) domain remains a critical bottleneck in OTA testing.

\begin{figure}[!t]
	\centering
	\includegraphics[width=0.48\textwidth]{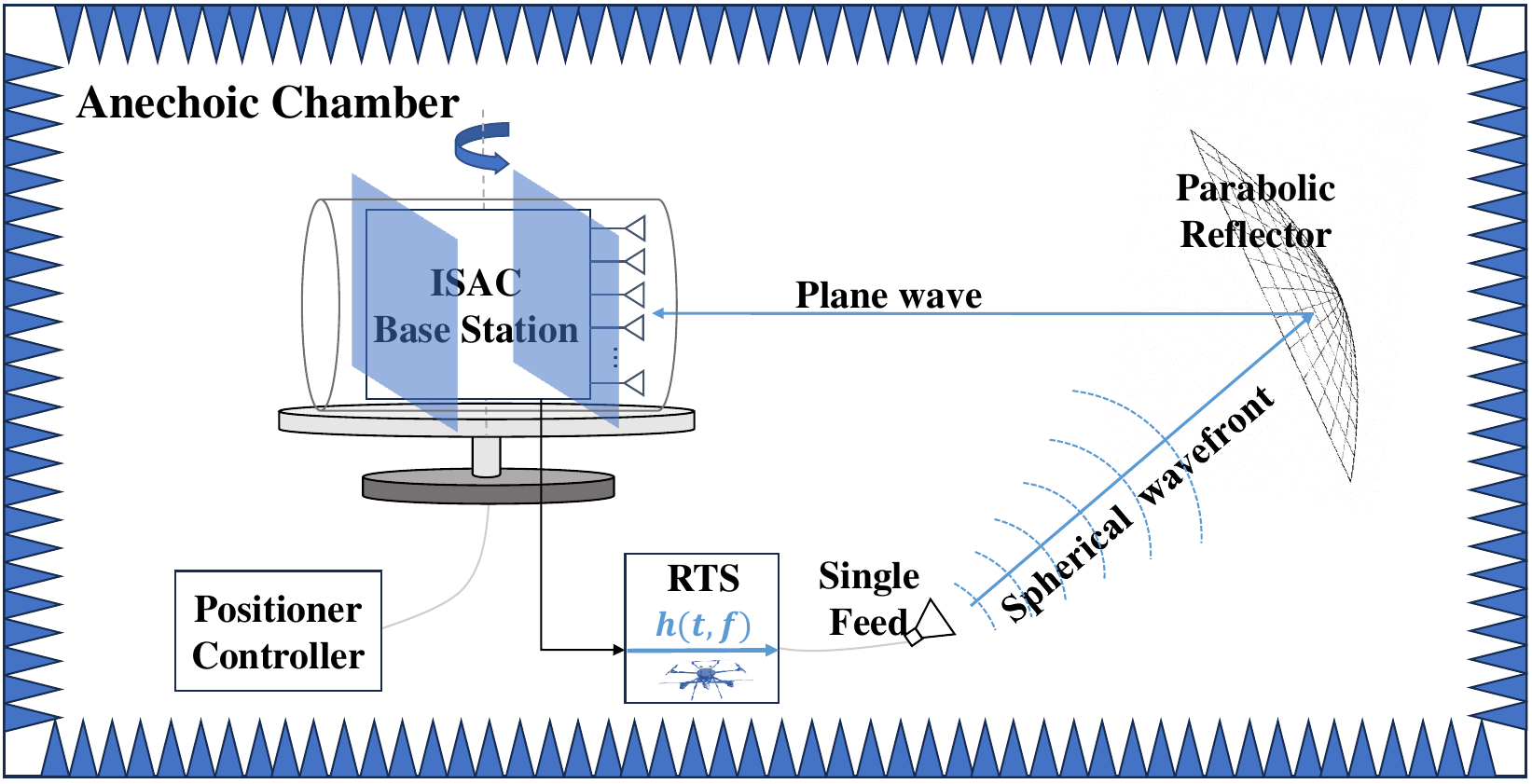}
	\caption{The framework of conventional BS test system.}
	\label{framework1}
\end{figure}

The current conventional BS test system based on CATR, as shown in Fig. \ref{framework1}, typically generates a single angular plane wave using a parabolic reflector with a single feed, necessitating the mechanical rotation of the BS under test to sweep different angles \cite{tancioni2019over}. However, it is incapable of generating simultaneous multi-angle wavefronts, nor can it emulate dynamic targets with time varying spatial profiles. Therefore, the primary challenge addressed in this work is to overcome the spatial limitations of conventional architectures, realizing the emulation of closely spaced targets with dynamically adjustable AoA without relying on mechanical movement. {To the best of our knowledge, there has been currently no reports of such system framework can flexibly emulate multiple angles plane waves for ISAC BS testing under FF conditions.}

\subsection{The Proposed Framework}
\begin{figure}[t]
	\centering
	\includegraphics[width=0.48\textwidth]{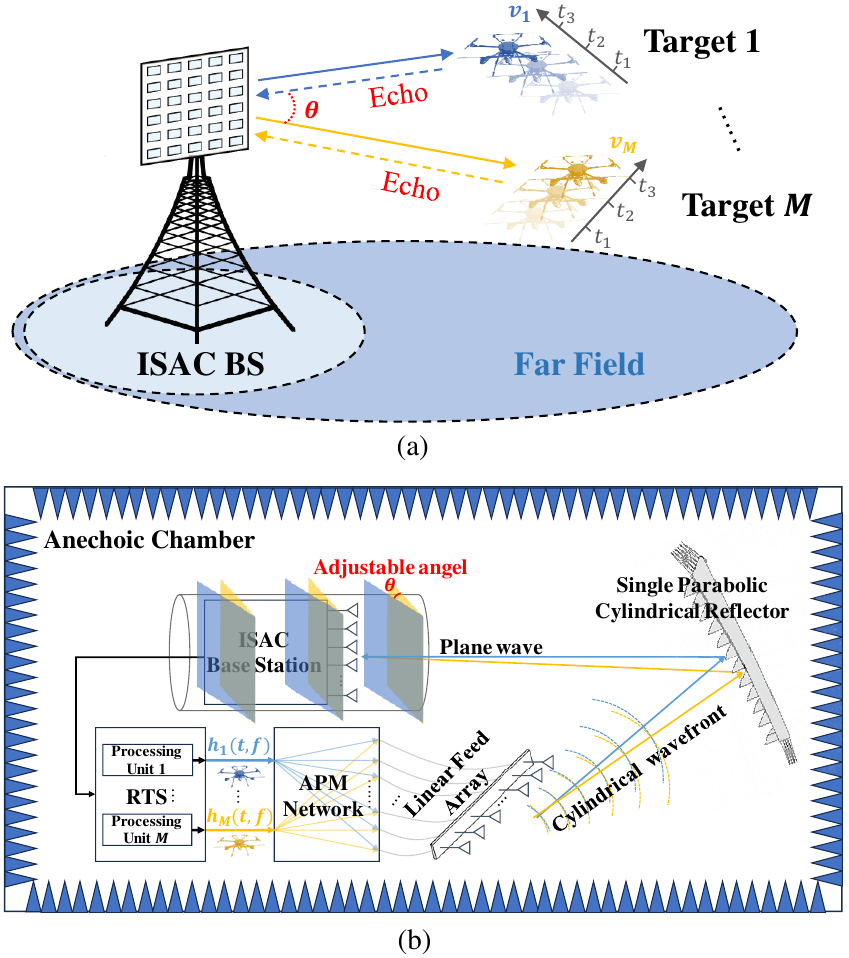}
	\caption{(a) The emulated multi-target scenario. (b) The framework of proposed ISAC BS test system.}
	\label{framework2}
\end{figure}

{The framework of our proposed ISAC BS test system is shown in Fig. \ref{framework2} (b), composed of a RTS, multi-port APM network, linear array feed, single parabolic cylindrical reflector (SPCR) and the ISAC BS under test.} {It utilizes a closed-loop architecture for hardware in the loop emulation. The ISAC BS loads the transmit signals to the RTS which includes multiple processing units. The RTS is connected to the APM network, which is linked to the linear array feed via cables.} The linear array feed is uniformly distributed in a linear arrangement, mounted on the antenna bracket, and directed towards the SPCR. The antenna  bracket is positioned at the focal point of the SPCR. The ISAC BS under test is placed in the QZ area, which is covered by the plane wave from different incident angles generated by the cylindrical wave reflected from the SPCR. {The fundamental operational principle of our proposed system is identical to that of widely used commercial RTS. However, while commercial products are typically utilized in conducted (cabled) setups and are limited to emulating target RCS, Doppler, and range, our proposed SPCCR-based framework advances this by enabling OTA evaluation with the added dimension of highly controllable spatial AoA.}

The functionality of key modules are detailed as below:
\begin{itemize} 
	\item [1.]
	\textbf{RTS:} Use for emulating the FF point target characterized by RCS, range and Doppler, and load these emulated target responses on the transmit signal from ISAC BS under test. There includes multiple processing units inside the RTS, which can be used for emulating multiple targets in above mentioned dimensions, while the spatial dimension of targets will be supported by following APM network and SPCCR. In recent studies \cite{10887342}, we have demonstrated that the channel emulators (CE) can be used for multiple sensing targets emulation. 
	
	\item [2.]
	\textbf{APM network:} Use for adjusting the amplitude and phase for each feed in the linear array and combine the emulated target responses from the RTS.  {The APM network in our proposed framework has multiple input ports. We can simultaneously load multiple RTS signals emulating multiple targets to the APM network, the complex amplitude and phase weights for each specific target are linearly superimposed. By optimizing the amplitude and phase weights of the linear feed array, plane waves with different incident azimuth angles can be synthesized within the QZ. By electronically updating these weights through the APM network, the target AoA can be dynamically adjusted without mechanical movement.} 
	
	\item [3.]
	\textbf{SPCR:} Use for reflecting the cylindrical wavefront produced by linear array feed to plane wave. {Unlike conventional parabolic reflector which possess 3D rotational symmetry, the SPCR is parabolic in the vertical plane and flat in the horizontal plane, converting spherical waves into cylindrical waves. The cylindrical wave from the linear array feed is transformed into a plane wave after phase correction by the SPCR. The QZ in the SPCCR can approximate plane waves in the vertical direction (elevation plane), attributed to the correction of the SPCR. Consequently, the optimization of the incident plane wave direction is inherently constrained to the azimuth domain (horizontal plane), rendering the elevation angle non-optimizable via the reflector's geometry.}
	
	\item [4.]
	\textbf{ISAC BS:} {As the DUT, it loads transmit signal to the RTS, and receives the echo signal with different incident angles generated by the linear array feed. The ISAC BS can transmit any arbitrary or dynamic waveform, and our system will accurately echo it back, which also works for radar.}
	
\end{itemize}

The proposed testing scheme offers several distinct advantages over existing methodologies. {First, by introducing a novel SPCCR architecture integrated with a multi-feed amplitude phase controlled linear array, APM network with multiple input ports and RTS, the system supports simultaneously emulating multiple closely-spaced spatial targets with a $\pm 10^\circ$ AoA coverage in the azimuth plane. It is worth noting that $\pm 10^\circ$ AoA limitation in our current setup is a practical engineering tradeoff determined by the width of QZ and the angle of incident waves, which will be demonstrated in later.} Furthermore, it overcomes the limitation of insufficient port availability in commercial RTSs, providing a scalable solution for validating ISAC BSs. Second, by leveraging the fundamental architecture of the CATR, the proposed test system inherits the intrinsic advantages of this mature and well-established technology, which offers ultra-wide system bandwidth, covering all frequency bands of interest for ISAC applications by interchanging the linear array feed and the APM network without altering the main reflector, while simultaneously providing a sufficiently large QZ to accommodate electrically large antenna apertures. Third, the system significantly enhances testing efficiency for radiation pattern measurements. In contrast to traditional setups relying on mechanical positioning, where measuring a full $360^\circ$ pattern with $1^\circ$ resolution typically necessitates $360$ discrete mechanical rotations. Our approach synthesizes plane waves with arbitrary incident angles solely through electronic control of the linear feed array and the APM network. By leveraging an electronic scanning range of $\pm 10^\circ$, the proposed system requires only $18$ mechanical rotations (one every $20^\circ$) to cover the entire azimuth plane. Within each mechanical step, fine angular sweeping is handled instantaneously by the APM, thereby eliminating mechanical latency for the vast majority of data points and facilitating rapid dynamic testing. Finally, the proposed system provide a highly versatile and cost-effective platform for future 6G testing campaigns.

\subsection{{Discussion}}

\begin{itemize} 
	\item [1.] 
	{\textbf{Support for wideband testing}}
	 
	{The proposed framework is applicable to wideband testing. The amplitude and phase weights for linear feed array are optimized at a selected center frequency and remain effective over a finite bandwidth around that frequency, which will be discussed in the following section. The practically supported instantaneous bandwidth is therefore determined by the frequency stability of the RTS, APM network, linear feed array, and associated RF components.  Although the APM network exhibits constrained instantaneous bandwidth, its carrier frequency can be freely tuned to support operations across a wide system frequency range.}
	
	\item [2.]
	{\textbf{Support for multiple frequency bands operation}}
	
	{The proposed framework adopts a modular architecture for operation at different frequency bands. The SPCR can be retained, whereas the APM networks and linear feed array are replaced according to the target frequency bands. The corresponding RTS and RF components can also be reconfigured when required.} 
	
	\item [3.]
	{\textbf{Support for multiple targets emulation}}

	\begin{itemize} 
		\item [3.1]	
		
		{The multi-port, and amplitude-phase controlled architecture is the key hardware basis that enables the proposed system to emulate multiple targets. Unlike conventional phased array feed networks that typically feature only a single input port and synthesize only one spatial field distribution at a time, the proposed APM network is engineered with multiple input ports. This multi-port topology allows multiple independent RF signals to be injected into the APM network simultaneously. Each input signal can be connected to an independent RTS processing unit and can therefore carry an individually configured target response, including range, Doppler, and RCS information. Meanwhile, the corresponding amplitude and phase weights across the linear feed array determine the spatial AoA profile of the target. By updating the pre-configured amplitude/phase weighting states of the APM network, the steering angle of the generated plane wave can be dynamically changed with negligible hardware latency.} 
		
		\item [3.2]
		
		{More importantly, the proposed system has the capability of emulating closely spaced spatial targets with small angular separations, which is essential for evaluating the practical angular resolution limit of ISAC BSs. When multiple targets arrive from adjacent directions, their spatial signatures become highly correlated, making reliable detection and separation particularly challenging. Reproducing such scenarios in a controlled and repeatable OTA environment enables rigorous validation of high-resolution DoA estimation, digital beamforming, and multi-target discrimination under realistic RF-chain impairments, array calibration errors, and nonideal QZ conditions. This capability is therefore critical for determining whether future large aperture ISAC BSs can preserve their theoretical angular resolution in practical deployments.}
		
		\item [3.2]
		{However, the current implementation is limited to the emulation of multiple closely spaced targets within a finite AoA range in the azimuth plane. The achievable angular coverage depends on the tradeoff between the required QZ size and the acceptable amplitude and phase deviations within the QZ, as quantitatively investigated in the subsequent sections. To extend the evaluation to the elevation domain, the DUT can be mechanically rotated by 90 degree on a turntable, allowing the same electronic synthesis method to be applied sequentially in both angular domains.}
		
	\end{itemize}
	\item [4.]
	{\textbf{System cost}}
	
	{The proposed system incurs a higher initial cost than a conventional single-feed CATR due to the additional multi-port APM network and linear feed array. Recent years have witnessed remarkable progress in cutting down the costs of amplitude and phase control matrices, owing to the extensive deployment and mature technology of APM systems even at millimeter-wave frequencies. Moreover, the SPCR can be reused across different frequency bands, making the proposed framework a practical and scalable solution for multi-band and multi-target OTA testing.}

\end{itemize}

\section{Multiple Targets Emulation Algorithm}
Consider a test scenario consisting of $M$ distinct targets shown in Fig. \ref{framework2} (a), to evaluate the sensing capability of the ISAC BS. {In this scenario, the RTS with $M$ processing units are employed, where an APM network establishes a fully connected mapping between the $M$ processing units and a linear array composed of $L$ antenna elements. The output signal matrix $\boldsymbol{h} \in \mathbb{C}^{M \times 1} $ from the RTS, corresponding to the $M$ processing units, can be expressed as:}
\begin{equation}
	{\boldsymbol{h} = [ {h}_1(t,f), {h}_2(t,f), ... ,{h}_M(t,f) ]^{T},}
	\label{RTS}
\end{equation}
{where $h_m(t,f)$ denotes the emulated target response for the $m\text{-th}$ processing unit, given by:}
\begin{equation}
	{{h}_m (t,f) = G_m(t) \cdot \text{exp}(j 2 \pi \nu_m(t) t) \cdot \text{exp}(j 2 \pi f \tau_m(t) ), }
	\label{RTS-CFR}
\end{equation}
{where $m = 1,...,M$. $G_m(t)$ is the channel gain for the $m\text{-th}$ emulated target, determined by path loss and RCS. $\nu_m(t)$ and $\tau_m(t)$ are the Doppler shift and delay of the $m\text{-th}$ emulated target at time $t$, which are determined by the target velocity and range, respectively. The APM is used to adjust the amplitude and phase of the linear array feed, which splits these RTS signals for the $L$ feeds. The corresponding amplitude and phase coefficients $\boldsymbol{e}_m = \{ e_m^l(t,f) \} \in \mathbb{C}^{1 \times L} $ associated with the $m\text{-th}$ emulated target for the $l\text{-th}$ feed can be given by}
\begin{equation}
	{{e}_m^l (t,f) = A_m^l(t) \cdot \text{exp}(j \phi_m^l(t,f)),}
	\label{APM}
\end{equation}
where $A_m^l(t)$ and $\phi_m^l(t,f)$ represent the amplitude and phase for the $l\text{-th}$ feed at time $t$, respectively. Therefore, the excitation weight matrix $\mathbf{W} \in \mathbb{C}^{M \times L} $ for linear array feed combined the output of all RTSs and APM can be expressed as:
\begin{equation}
	\mathbf{W} = [ \boldsymbol{w}_1; \boldsymbol{w}_2 ; ... ;\boldsymbol{w}_M  ],
	\label{Feed}
\end{equation}
where the weight of feeds $\boldsymbol{w}_m = \{ w_m^l(t,f) \} \in \mathbb{C}^{1 \times L}  $ for $m\text{-th} $ emulated target is defined as:
\begin{equation}
	{{w}_m^l (t,f) = {e}_m^l (t,f) \cdot  {h}_m (t,f).}
	\label{Feed1}
\end{equation}

The electric field $\mathbf{G} \in \mathbb{C}^{M \times N} $ generated by linear array feed in the QZ can be efficiently computed by the following formula:
\begin{equation}
	{\mathbf{G} = \mathbf{W} \mathbf{S},}
	\label{Optimization}
\end{equation}
where $\mathbf{S} = [\boldsymbol{s}_1; \boldsymbol{s}_2;...;\boldsymbol{s}_L ] \in \mathbb{C}^{L \times N}$ is the transfer function matrix between each antenna feed and sampling points $N$ in QZ. $\boldsymbol{s}_L \in \mathbb{C}^{1 \times N}$ denotes the transfer function vector for the $L\text{-th}$ feed. {It should be noted that $\mathbf{G}$ specifically denotes the dominant co-polarized component of the electric field. Because the cross-polarized components in this SPCCR architecture are generally suppressed below -30 dB within the QZ \cite{yang2024enhancing}, they are considered negligible. Consequently, the subsequent amplitude and phase optimizations, as well as all field distributions plotted in Sections III and IV, are evaluated exclusively for this selected co-polarized component.}

Owing to the use of a SPCR, the optimization can only be performed in the azimuthal (horizontal) dimension. As a result, plane waves with different azimuth angles can be generated and directed toward the QZ. The field sampling within the QZ is restricted to the horizontal and vertical centerlines. Specifically, a total of $N = N_h + N_v$ sampling points are considered, where $N_h$ and $N_v$ denote the numbers of sampling points along the horizontal and vertical centerlines, respectively. This sampling strategy captures the dominant field variations while significantly reducing the computational complexity. To simplify the subsequent optimization procedure, we consider only the horizontal dimension electric field which can by expanded as:
\begin{equation}
	\mathbf{G} = [\boldsymbol{g}_1 (x,\theta_1); \boldsymbol{g}_2 (x,\theta_2); ...;\boldsymbol{g}_M (x,\theta_M) ],
	\label{Optimization1}
\end{equation}
where $\boldsymbol{g}_m (x,\theta_M) \in \mathbb{C}^{1 \times N}$ represents the generated electric field of the $m\text{-th} $ emulated target in the $x\text{-}$direction with the incident angle $\theta_m$. It can be further expressed as:
\begin{equation}
	{\boldsymbol{g}_m (x,\theta_m) = \boldsymbol{g}_0 \cdot \text{exp}(-j k x \text{cos}\theta_m),}
	\label{G}
\end{equation}
where $\boldsymbol{g}_0 \in \mathbb{C}^{1 \times N}$ is a complex amplitude vector that determines the polarization, magnitude, and initial phase of the electric field \cite{lindell1991methods}. $k$ is the wavenumber.

\begin{figure}[t]
	\centering
	\includegraphics[width=0.48\textwidth]{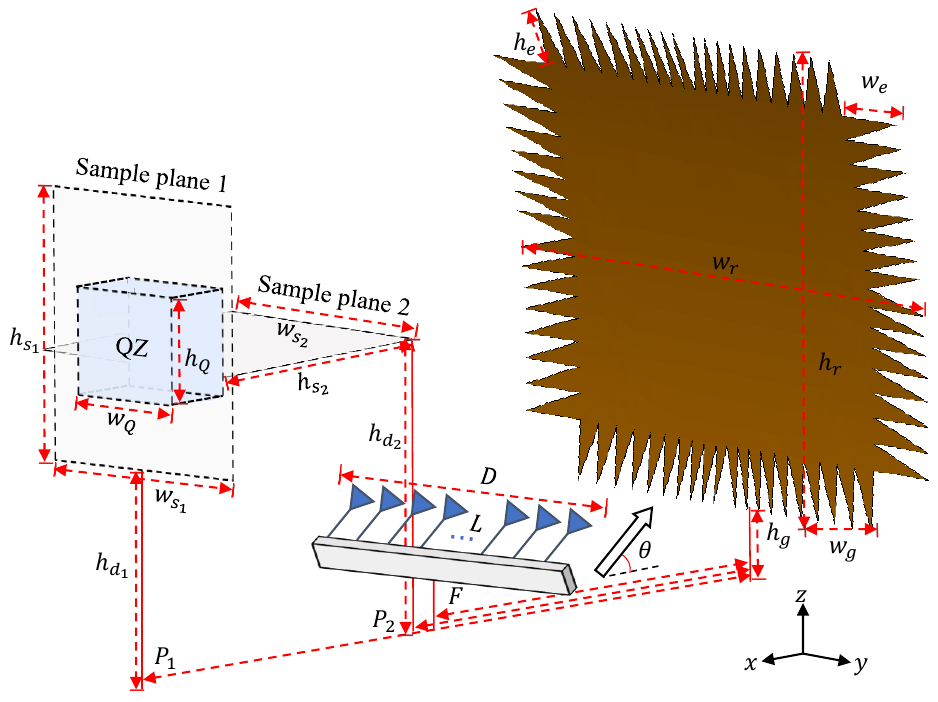}
	\caption{{The schematic of the SPCCR based on linear array feed.}}
	\label{SPCR}
\end{figure}

\begin{table}[t]
	\centering
	\caption{{The Parameters for the SPCCR}}
	\label{tab:spccr_params}
	\begin{tabular}{ccc}
		\toprule
		\textbf{Parameter} & \textbf{Meaning} & \textbf{Value} \\
		\midrule
		$h_e$ & height of the edge teeth & 0.47 m \\
		$w_e$ & width of the edge teeth & 0.54 m \\
		$h_r$ & height of reflector & 3.66 m \\
		$w_r$ & width of reflector & 4 m \\
		$h_g$ & \begin{tabular}{@{}c@{}}distance from the lower edge \\ of reflector to the ground\end{tabular} & 0.42 m \\
		$w_g$ & \begin{tabular}{@{}c@{}}horizontal distance between the upper \\ and lower edges of the reflector\end{tabular} & 0.99 m \\
		$h_{s_1}$ & height of sample plane 1 & 3 m \\
		$w_{s_1}$ & width of sample plane 1 & 4 m \\
		$h_{s_2}$ & height of sample plane 2 & 3 m \\
		$w_{s_2}$ & width of sample plane 2 & 4 m \\
		$h_{Q}$ & height of QZ & 1 m \\
		$w_{Q}$ & width of QZ & 2 m \\
		$h_{d_1}$ & distance of sample plane 1 to the ground & 1 m \\
		$h_{d_2}$ & distance of sample plane 2 to the ground & 2.27 m  \\
		$P_{1}$ & distance of sample plane 1 to the origin & 7 m \\
		$P_{2}$ & distance of sample plane 2 to the origin & 4.32 m \\
		$F$ & focal length of the reflector & 4 m \\
		$D$ & length of linear array & 5.2 m \\
		$L$ & Number of linear array feed & 44 \\
		$\theta$ & propagation direction of waves & $32^{\circ}$ \\
		\bottomrule
	\end{tabular}
\end{table}

By optimizing the excitation weights $\mathbf{W}$ of the linear feed, plane waves with different azimuth angles can be generated in the QZ for evaluating the target sensing capability of the ISAC BS. Since plane waves with different incident angles correspond to different field distributions in the QZ, the optimization problem is solved independently for each target angle $\theta_m$, $m=1,\ldots,M$. Therefore, for the $m$-th emulated target, the excitation weight vector
$\boldsymbol{w}_m$ defined in \eqref{Feed1} is obtained by solving the
following constrained least-squares (LS) problem \cite{finlayson1997constrained}:
\begin{equation}
	\begin{aligned}
		\min_{\boldsymbol{w}_m} \quad
		& \left\| \boldsymbol{w}_m \mathbf{S}
		- \boldsymbol{g}_m(x,\theta_m) \right\|_2^2 \\
		\text{s.t.}\quad
		& \|\boldsymbol{w}_m\|_2 \le \eta \,
		\|\boldsymbol{w}_m^{\mathrm{LS}}\|_2 ,
	\end{aligned}
	\label{LS}
\end{equation}
{where $\boldsymbol{w}_m^{\mathrm{LS}}$ denotes the unconstrained LS solution, and $0<\eta<1$ is a constraint coefficient controlling the allowable excitation power of the linear feed. The power outside the QZ can be controlled by adjusting the value of $\eta$. We can reduce the value of $\eta$ to make power concentrate more in QZ and lead to less spurious radiation.} Substituting $\boldsymbol{w}_m = h_m \boldsymbol{e}_m$ into \eqref{LS}, the optimization problem can be equivalently expressed as
\begin{equation}
	\begin{aligned}
		\min_{\boldsymbol{e}_m} \quad
		& \left\| (h_m \boldsymbol{e}_m)\mathbf{S}
		- \boldsymbol{g}_m(x,\theta_m) \right\|_2^2 \\
		\text{s.t.}\quad
		& \|(h_m \boldsymbol{e}_m)\|_2
		\le \eta \,
		\|\boldsymbol{w}_m^{\mathrm{LS}}\|_2 , \\
		& 0 \le A_m^l \le A_{\max}, \quad
		\phi_m^l \in [-\pi,\pi), \; l=1,\ldots,L ,
	\end{aligned}
	\label{LS-APM}
\end{equation}
where $A_{\max}$ represents the maximum controllable amplitude provided by the APM for each antenna feed. {It is important to clarify the handling of wideband signals in the optimization problem described in \ref{LS-APM}, the phase coefficients $\phi_m^l$ are optimized at the center frequency $f_c$ of the system. Because practical APM apply frequency-independent control states. It is physically unfeasible to apply a continuously varying, frequency-dependent phase weight across a wideband pulse at the hardware level. And the geometry of the SPCR is frequency-independent, and the radiation characteristics of the feed elements vary relatively slowly within the designated sub-bands, the optimal weights derived at the center frequency remain highly robust across a considerable bandwidth.}

\section{Simulation and Discussion}
In this section, we conduct extensive simulation in FEKO with physical optics (PO) method \cite{asvestas1980physical} to validate the ability of emulating closely spaced target in azimuth domain for our proposed ISAC BS test system.

\begin{figure*}[t]
	\centering
	\includegraphics[width=1\textwidth]{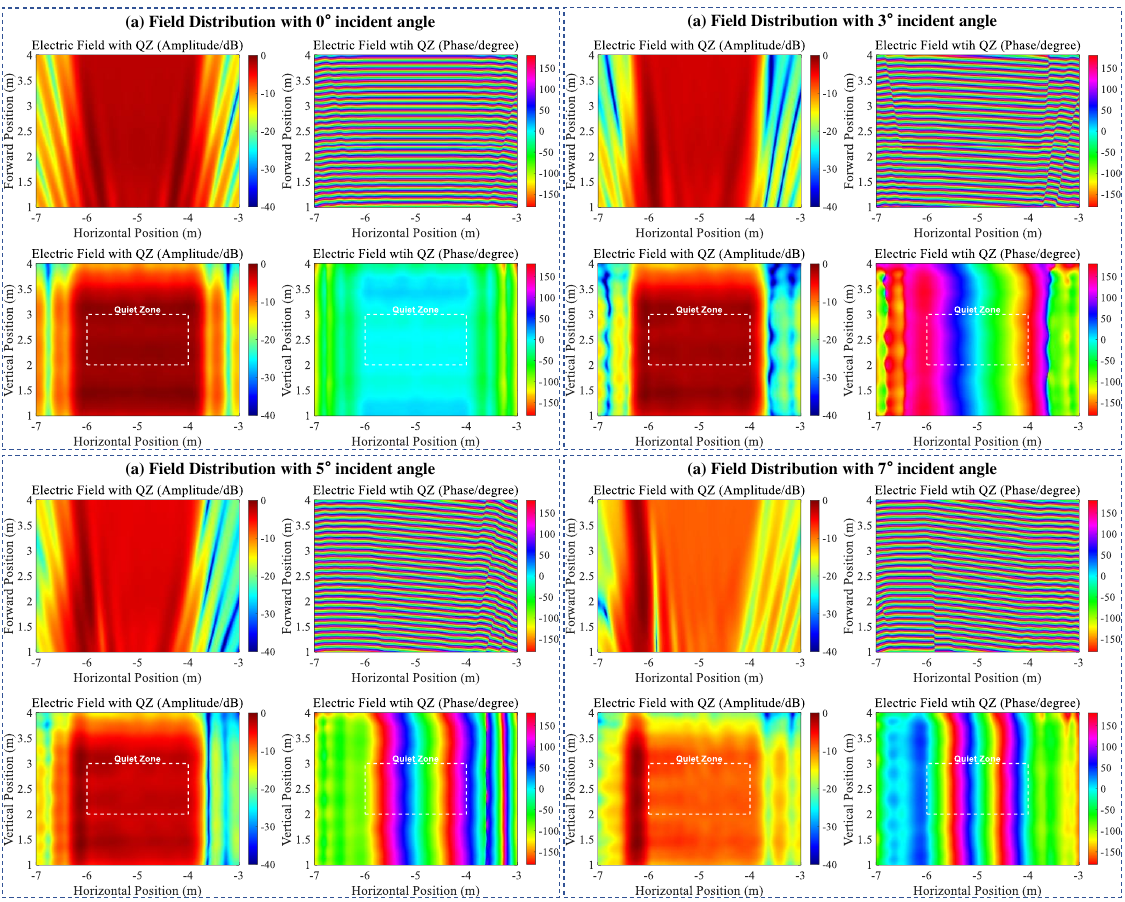}
	\caption{Simulated electric field distributions within the target QZ for various incident azimuth angles. The subfigures illustrate the amplitude (left) and phase (right) distributions on two sampled planes for incident angles of (a) $\theta_1 = 0^\circ$, (b) $\theta_2 = 3^\circ$, (c) $\theta_3 = 5^\circ$, and (d) $\theta_4 = 7^\circ$, respectively. The white dashed rectangles indicate the designated QZ region.}
	\label{E}
\end{figure*}

\subsection{Simulation Setup}
We adopted the linear array feed SPCCR in \cite{yang2024enhancing}, with only a brief overview presented here. {The schematic of the SPCCR is shown in Fig. \ref{SPCR}, and the detailed parameters are all concluded in Table \ref{tab:spccr_params}. The linear array utilized in this simulation is a uniform linear array (ULA) positioned at the focal point, which consists of 44 radiating elements with operating frequencies set to 2.6 GHz, where the radiation pattern for each feed is modeled using the imported data of a corrugated horn antenna. These elements are distributed in an equidistant arrangement with a spacing of $1.4 \lambda$. The first observation region is set 3 m behind the linear array, defined as a planar surface spanning 4 m horizontally and 3 m vertically. The field distribution within this region is discretized into a sampling grid of 401 $\times$ 301 points. The second observation region is set perpendicular to the first observation region, configured as a horizontal plane parallel to the ground. This surface spans 4 m in the horizontal direction and 3 m in the forward propagation direction, and is similarly discretized into a sampling grid of $401 \times 301$ points.}

\begin{figure*}[t]
	\centering
	\includegraphics[width=1\textwidth]{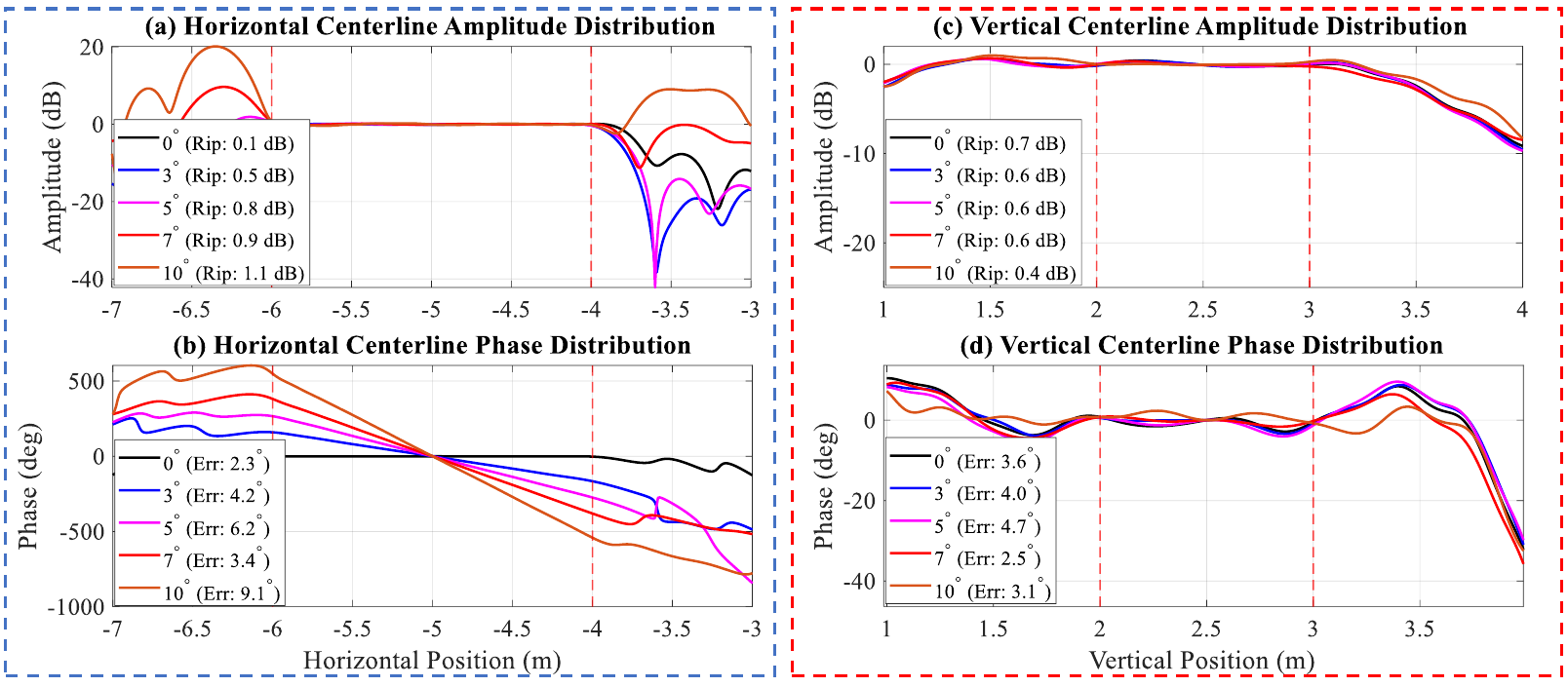}
	\caption{(a) The amplitude distribution along the horizontal centerline. (b) The phase distribution along the horizontal centerline. (c) The amplitude distribution along the vertical centerline. (d) The phase distribution along the vertical centerline. }
	\label{ripple}
\end{figure*}

The transfer function from each feed element to the first sampling plane was collected by exciting the array elements sequentially. A target QZ with dimensions of $2\text{ m} \times 1\text{ m}$ (horizontal $\times$ vertical) was defined. Specifically, the field distributions along the horizontal and vertical centerlines within this QZ in first observation region were selected as the optimization objectives. Based on the acquired transfer functions, the optimization of excitation coefficients was performed independently for each specific target angle to minimize field deviation. Here, we emulated five distinct targets corresponding to incident azimuth angles of $0^\circ$, $3^\circ$, $5^\circ$, $7^\circ$ and $10^\circ$, respectively.

\subsection{Simulation Results}
The simulated electric field distribution in two sampled planes within the target QZ under multiple incident azimuth angles are shown in Fig. \ref{E}, illustrating both the amplitude and phase of the electric field, respectively. Specifically, the white dashed rectangles indicate the QZ area. We can observe that with the increase of the incident angle, while the amplitude flatness is well-preserved within the target QZ, the amplitude levels outside the QZ boundary increase significantly, ultimately exceeding the amplitude observed within the target QZ area. However, it is an unavoidable trade-off associated with wide incident angle, which cannot be fully mitigated by optimizing the feed excitations. {Specifically, this unavoidable trade-off encompasses a multi-dimensional physical balance among the QZ synthesis quality, the out-of-zone spillover power, and the effective QZ size. Generating plane waves at wider oblique angles requires steeper phase gradients across the linear feed array, which inherently demands significantly higher excitation power. In our optimization, the parameter $\eta$ defined in (\ref{LS}) actively controls this balance by restricting the $L_2\text{-}$norm of the excitation weights. While an unconstrained optimization (i.e., a larger $\eta$) could yield a nearly perfectly flat QZ, it forces highly oscillatory feed excitations, resulting in severe out-of-zone spillover that would exceed practical hardware amplifier limits.} 
\begin{figure}[!t]
	\centering
	\includegraphics[width=0.5\textwidth]{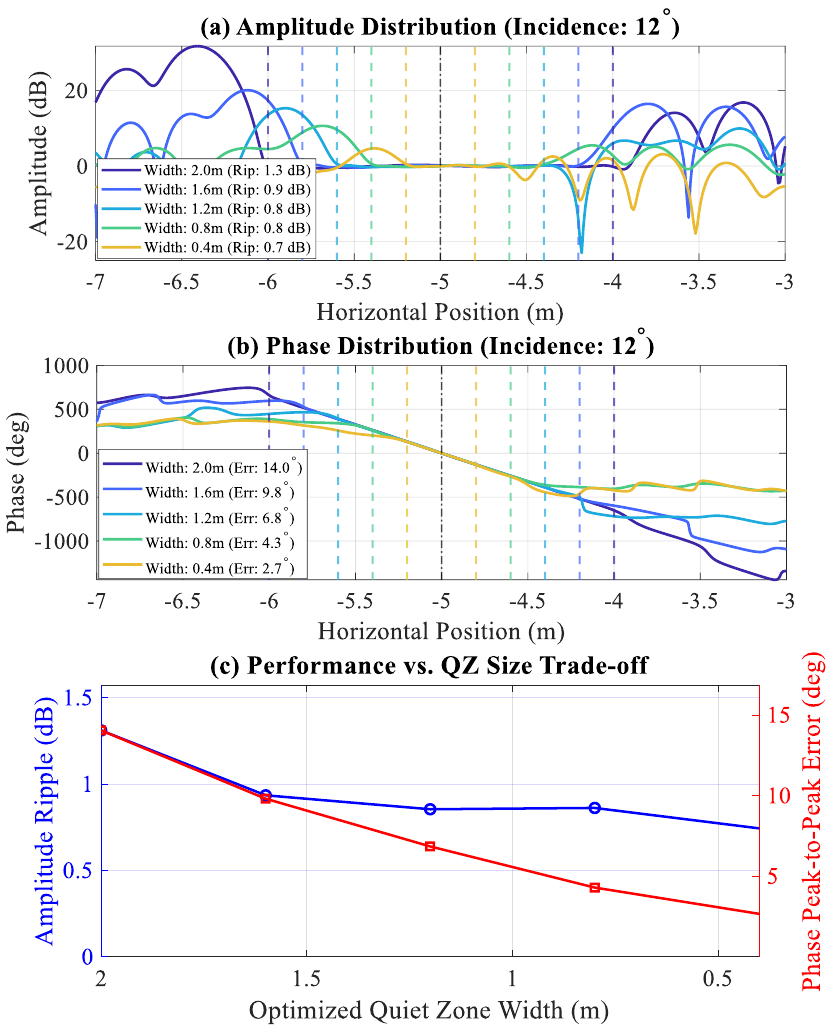}
	\caption{The (a) amplitude and (b) phase distribution along the horizontal centerline with the $12^{\circ}$ incident azimuth angle under different QZ widths. (c) Trade-off analysis between QZ width and QZ performance at $12^\circ$ incident angle. }
	\label{ripple_trend}
\end{figure}

To further evaluate the QZ quality, the amplitude and phase distributions along these principal centerlines are plotted in Fig. \ref{ripple}. The amplitude ripple is defined as the peak-to-peak variation of the normalized electric field amplitude within the target QZ. Correspondingly, the phase error is quantified as the peak-to-peak deviation of the synthesized phase profile relative to the theoretical linear phase gradient determined by the target incident angle. With the increase of incident angle, the amplitude ripple and phase error along horizontal centerline in QZ gradually increase, making the optimization of feed excitations increasingly challenging while maintaining a constant QZ width. On the other hand, we can find that the field distribution along the vertical centerline under different incident angles are almost the same, due to character of the SPCR that only exhibits parabolic curvature in the horizontal dimension. It indicates that the quality of QZ in vertical centerline cannot be optimized, it is dependent solely on the physical geometry of the reflector and the intrinsic radiation pattern of the feed elements.

Therefore, we prioritize the analysis of the field distribution along the horizontal centerline, as this dimension provides the necessary degrees of freedom to actively adjust the field profile through the optimization of excitations. To further investigate the relationship between incident angles and QZ width, we conducted simulations targeting a $12^\circ$ incident angle while varying the horizontal width of the target QZ, which is progressively reduced from the nominal 2 m to 0.4 m with a step of 0.4 m. The corresponding amplitude and phase distribution are shown in Fig. \ref{ripple_trend} (a) and (b). It can be observed that with the reduction of QZ width, the amplitude ripple and phase error both exhibit a decreasing trend, which is illustrated in Fig. \ref{ripple_trend} (c). This reveals an inherent trade-off between the effective QZ dimensions and the achievable angular range. Consequently, we can also emulate targets with wider incident angles while maintaining high field fidelity by sacrificing the QZ width.

\begin{figure*}[!t]
	\centering
	\includegraphics[width=1\textwidth]{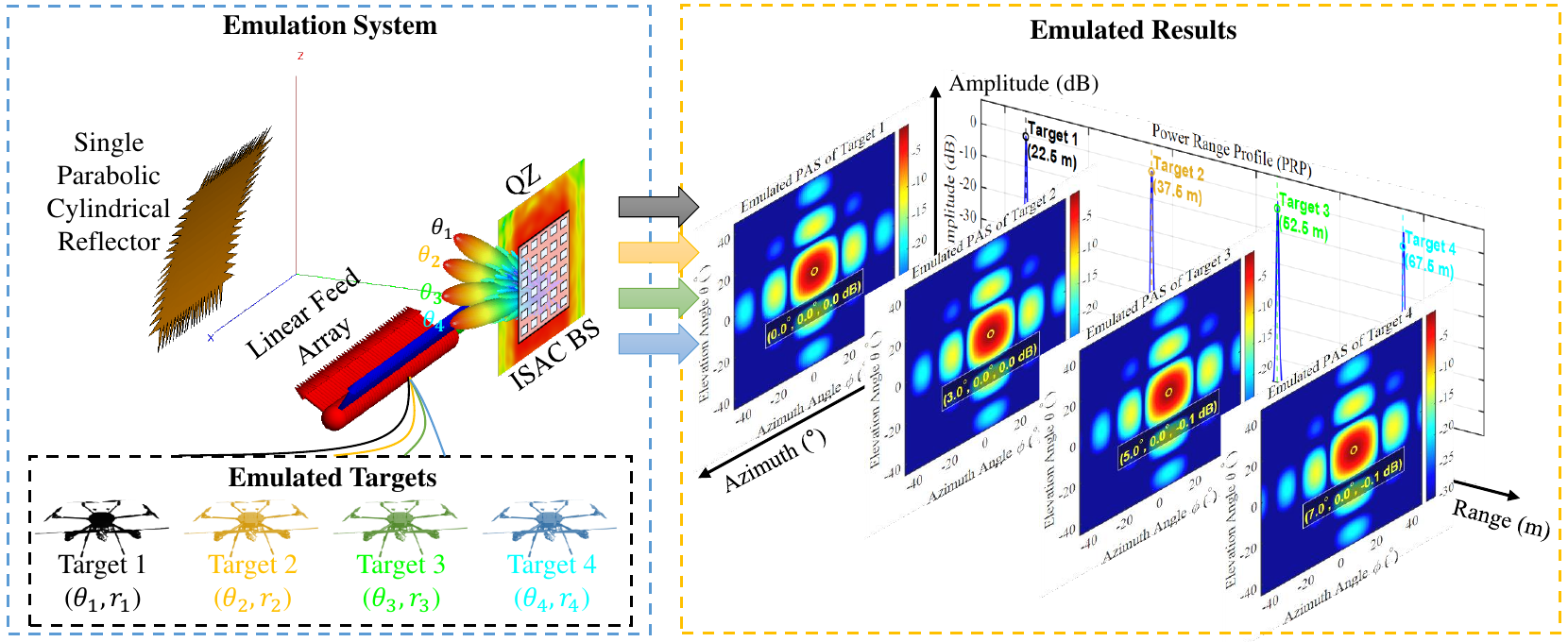}
	\caption{Illustration of the proposed SPCCR-based multi-target emulation framework: (Left) System configuration with the linear feed array and parabolic cylindrical reflector generating multiple incident wavefronts towards the ISAC BS in the QZ. (Right) Emulated results showing the power angle spectrum (PAS) and power range profile (PRP) for four distinct targets with varying angles and ranges. }
	\label{BF}
\end{figure*}

\begin{figure*}[!t]
	\centering
	\includegraphics[width=1\textwidth]{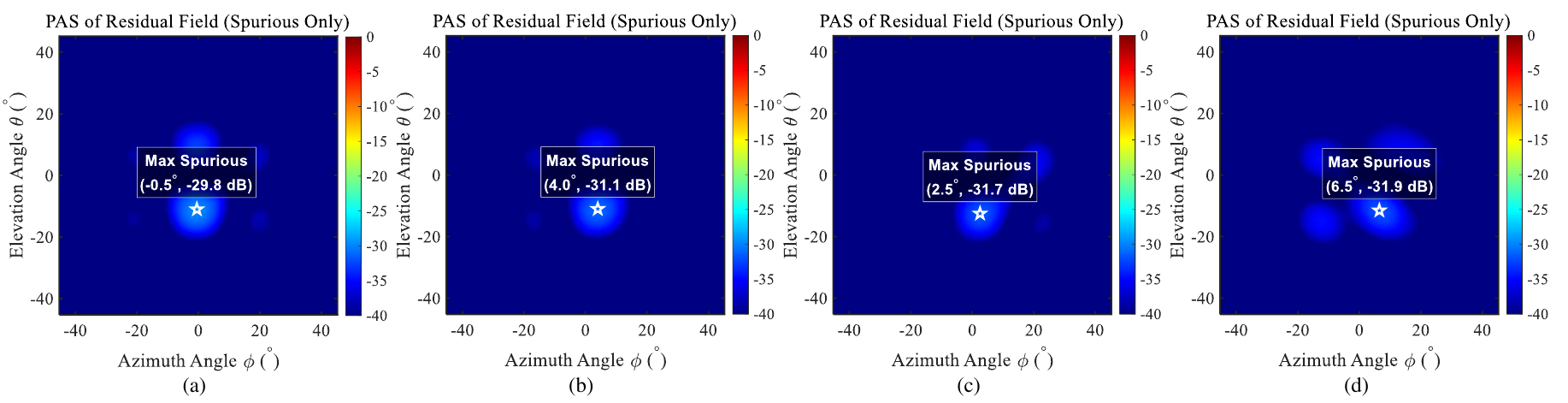}
	\caption{{The PAS of isolated residual field for various incident angles. (a) $\theta_1 = 0^\circ$, (b) $\theta_2 = 3^\circ$, (c) $\theta_3 = 5^\circ$, (d) $\theta_4 = 7^\circ$. }}
	\label{SFDR}
\end{figure*}

\subsection{Multiple Targets Emulation Results}

While RTS have achieved maturity in generating precise range and Doppler shifts, high-fidelity angular emulation, particularly for distinguishing closely spaced targets within a compact range environment, remains a critical challenge. Consequently, this section prioritizes the evaluation of the system's spatial accuracy and demonstrates its resolution capability in the joint angle-delay domain.

The illustration of proposed multi-target emulation framework is shown in Fig. \ref{BF}, where the system configuration with linear array feed generating multiple incident wavefronts towards the ISAC BS in the QZ is shown in the left of Fig. \ref{BF}. An $8 \times 8$ uniform planar array (UPA) was defined at the center of the optimized QZ to emulate a realistic ISAC BS. The array elements were spaced at half-wavelength ($\lambda/2$) intervals at $2.6\text{ GHz}$. To ensure physical validity, the received signals for each array element were extracted directly from the above FEKO simulation data grid using a nearest-neighbor search algorithm, rather than ideal interpolation. Subsequently, the classical beamforming algorithm was performed on these extracted signals to transform the spatial field distribution into the angular domain for spectral analysis, while other high resolution techniques can be used as well.

\textbf{1) Angular Fidelity Evaluation:}
Based on previous simulation, we configured a multi-target scenario involving four emulated targets where the configuration for each target is shown in Table. \ref{tab:target_parameters}. Besides the difference in angular domain, we allocate different ranges for each target. The power angular spectrums (PAS) based on these simulated data are shown in the right of Fig.\ref{BF}, where the emulated angular results perfect align with the theoretical value. When the incident angle are 5 and 7 degrees, there is a 0.1 dB deviation in the field amplitude, which can be ascribed to the slight degradation of amplitude flatness within the QZ at large oblique incident angles. {Moreover, we mathematically extracted the pure ideal plane wave component from the synthesized field sampled by the $8 \times 8$ array to obtain the residual field. Subsequently, perform beamforming on the residual field to isolate and quantify all spurious leakages. The PAS of isolated residual field are shown in Fig. \ref{SFDR}. While some optimization leakage exists, the strongest spurious spatial peak (the worst-case ghost target) is 29.8 dB lower than the intended target main lobe. In radar systems, a spurious free dynamic range (SFDR) exceeding 30 dB is generally sufficient to prevent false detection, ensuring that the out-of-zone field ripples do not translate into physical ghost targets at the ISAC BS receiver.}

\begin{table}[!t]
	\centering
	\caption{Comparison of Emulated and Ideal Target Parameters}
	\label{tab:target_parameters}
	\begin{tabular}{clcc}
		\toprule
		\textbf{Target Index} & \textbf{Parameter} & \textbf{Emulated} & \textbf{Theoretical} \\
		\midrule
		\multirow{3}{*}{Target 1 (T1)} 
		& Azimuth Angle $\theta_1$ ($^{\circ}$) & 0 & 0 \\
		& Range $r_1$ (m) & 22.5 & 22.5 \\
		& Amplitude (dB) & 0 & 0 \\
		\midrule
		\multirow{3}{*}{Target 2 (T2)} 
		& Azimuth Angle $\theta_2$ ($^{\circ}$) & 3 & 3 \\
		& Range $r_2$ (m) & 37.5 & 37.5 \\
		& Amplitude (dB) & 0 & 0 \\
		\midrule
		\multirow{3}{*}{Target 3 (T3)} 
		& Azimuth Angle $\theta_3$ ($^{\circ}$) & 5 & 5 \\
		& Range $r_3$ (m) & 52.5 & 52.5 \\
		& Amplitude (dB) & -0.1 & 0 \\
		\midrule
		\multirow{3}{*}{Target 4 (T4)} 
		& Azimuth Angle $\theta_4$ ($^{\circ}$) & 7 & 7 \\
		& Range $r_4$ (m) & 67.5 & 67.5 \\
		& Amplitude (dB) & -0.1 & 0 \\
		\bottomrule
	\end{tabular}
\end{table}

\begin{figure}[t]
	\centering
	\includegraphics[width=0.47\textwidth]{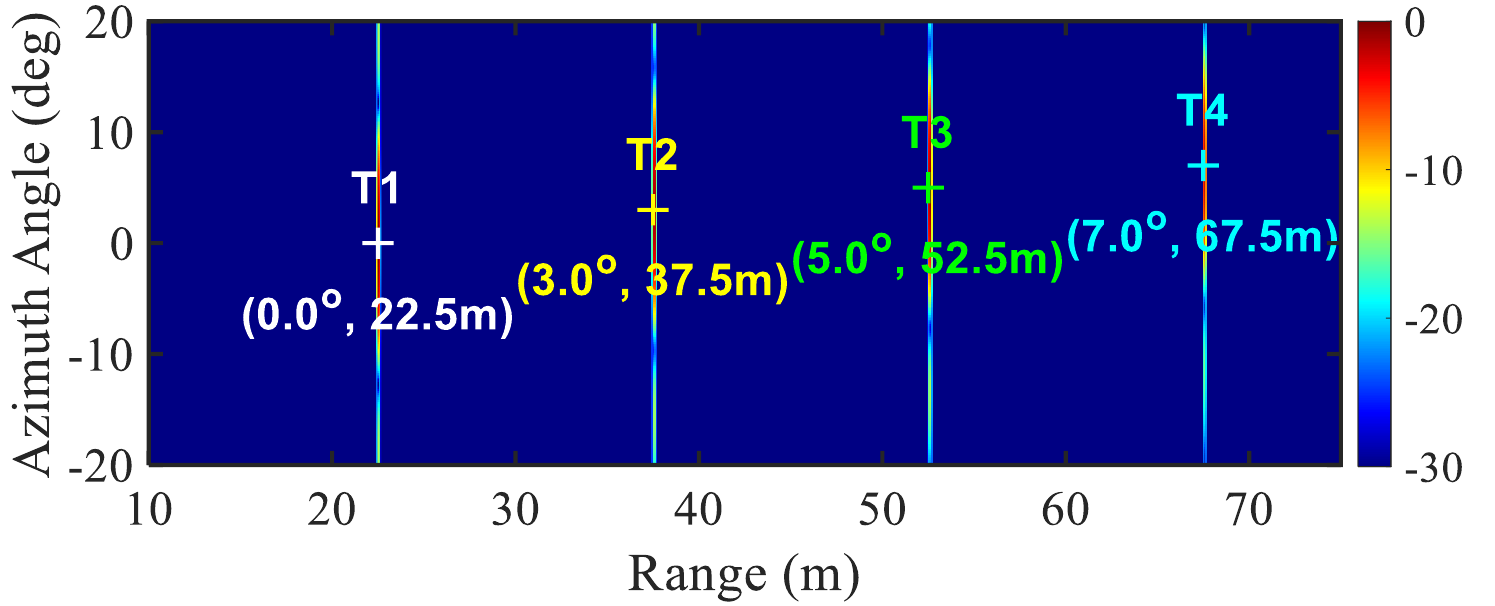}
	\caption{The PARP of a multi-target scenario involved four targets with different azimuth angles and ranges.}
	\label{delay}
\end{figure}

\textbf{2) Joint Angle-Range Evaluation:}
To further demonstrate the system's capability for multi-dimensional targets, we extended the analysis from the spatial domain to the joint angle-range domain. 

It is important to note that while the spatial field distribution was obtained from the aforementioned FEKO simulation at $2.6\text{ GHz}$, the wideband characteristics required for range resolution were synthesized via frequency-domain signal processing. Specifically, we extrapolated the single-frequency spatial data to a bandwidth of $2\text{ GHz}$ ($2-4\text{ GHz}$). This approach is valid given the frequency-independent geometric optics of the reflector, allowing us to emulate range without conducting computationally expensive wideband full-wave simulations. {Note that the synthesized 2 GHz wideband data used here is an idealized mathematical abstraction, utilized strictly to verify the joint angle-range signal processing algorithm without incurring prohibitive full-wave computational costs. This abstraction assumes frequency-independent feed characteristics. The rigorous validation of the system's true wideband physical performance, which intrinsically accounts for the frequency-dependent electrical array spacing and feed radiation patterns at the band edges is comprehensively demonstrated through the experimental measurements in Section V}. The resulting power angle range profile (PARP) is illustrated in Fig. \ref{delay}. The PARP clearly reveals four distinct energy concentrations. This result serves as a preliminary validation, confirming the feasibility of the proposed system for supporting multi-dimensional target emulation.

\begin{figure}[t]
	\centering
	\includegraphics[width=0.48\textwidth]{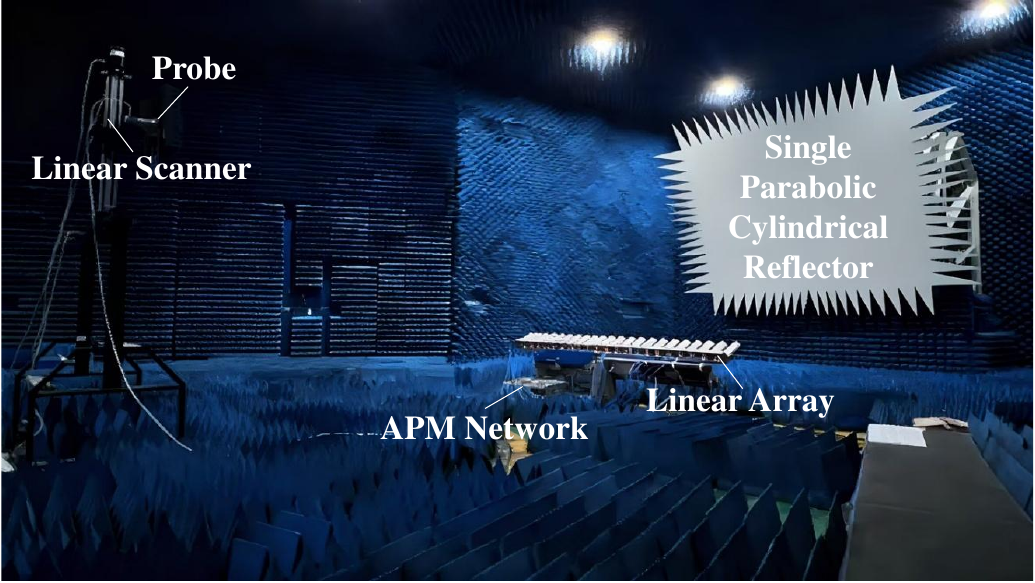}
	\caption{The measurement scenario based on SPCCR.}
	\label{measurement}
\end{figure}

\begin{figure*}[!t]
	\centering
	\includegraphics[width=1\textwidth]{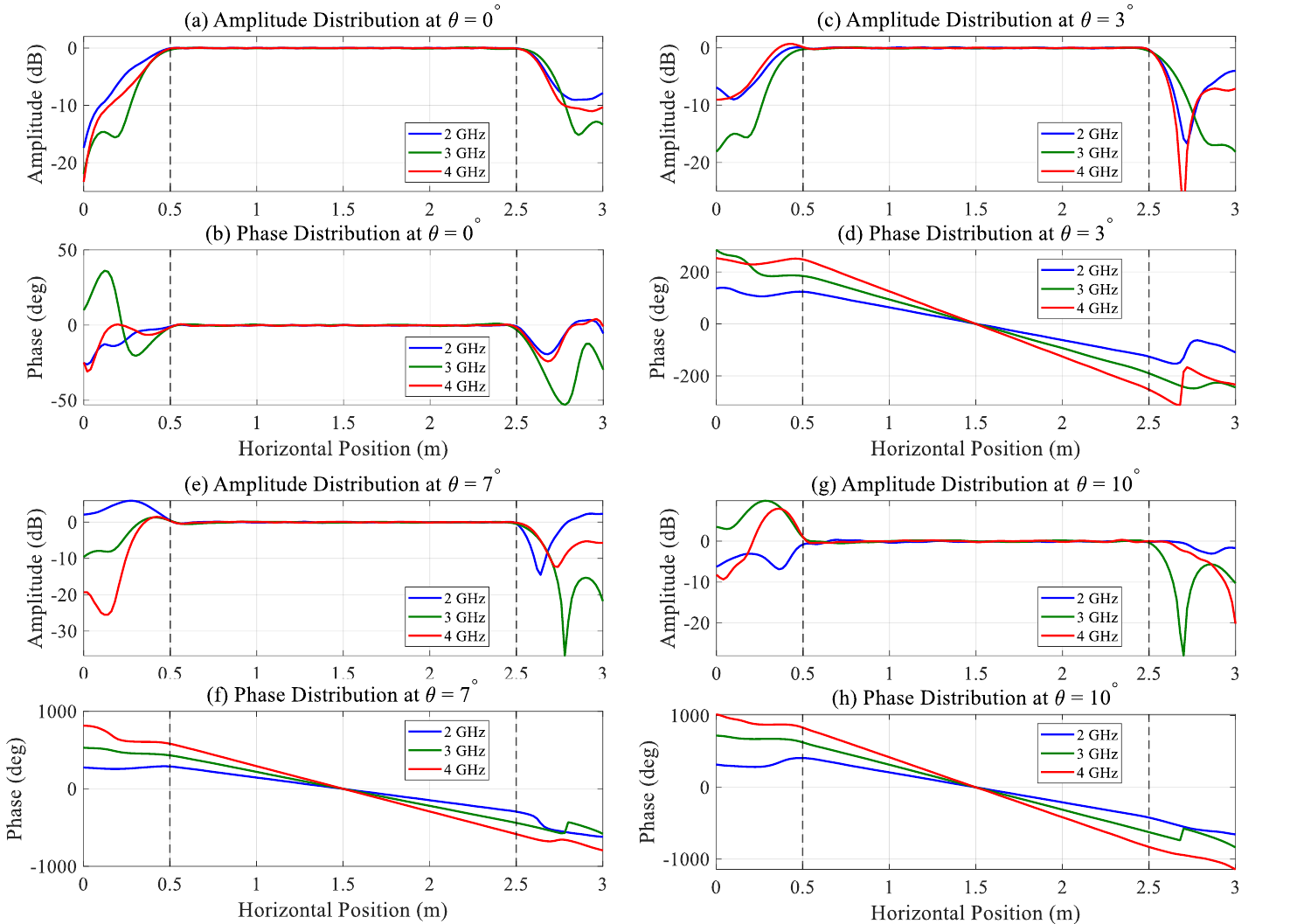}
	\caption{The horizontal field distribution at different frequencies under various incident azimuth angles based on measured transfer function.}
	\label{results}
\end{figure*}

\section{Experimental Validations}

This section aims to validate the practical feasibility of the proposed target emulation approach using measured transfer functions from the SPCCR testbed. Given that the full emulation process has been extensively detailed in the simulation section, this part focuses on demonstrating that the optimization method, when applied to real-world transfer functions, can successfully synthesize plane waves with different incident angles, thereby verifying the system's core capability in the spatial domain.

\subsection{Measurement campaign}
A comprehensive description of the measurement campaign is provided in \cite{yang2024enhancing}, with a brief introduction presented here. The measurement scenario based on SPCCR is shown in Fig. \ref{measurement}, where the entire system is placed in an anechoic chamber. The SPCR utilizes two-dimensional milling plates with high precision, designed according to the model in Section III. Owing to this high surface accuracy, the reflector supports an ultra-wide operating frequency range spanning from $1$ GHz to $110$ GHz. However, it is important to note that extending operation to higher frequency bands necessitates a denser linear feed array to avoid grating lobes, which leads to a significant increase in the required number of APM control channels. Due to the limited availability of high frequency APM hardware with sufficient channel capacity, the experimental validation in this work is restricted to the $2$--$4$ GHz range, although the system design allows for versatile target emulation across FR1, FR2, and FR3 bands by upgrading the feed and APM network. A linear array consisting of 44 uniformly spaced radiating elements is placed at the focal position of the SPCR. The array’s total length is 5.2 m, and the antennas are oriented towards the center of the reflector. Dual-polarized Vivaldi antennas are used. The probe antenna is mounted on a linear scanner to scan the horizontal centerline of the quiet zone. The quiet zone’s center is aligned with the reflector’s center, with a scanning range of 3 m and a step size of 200 mm. {Two 32-channel amplitude and phase control units, including a power distribution network, 8-bit phase shifters, and 9-bit attenuators, are connected to the array elements for precise control. Our current spatial target emulation and optimization are designed and evaluated exclusively for the primary co-polarized component.} A vector network analyzer (VNA) is used to measure the frequency response, with frequency points ranging from 2 GHz to 4 GHz.

\subsection{Measurement results}
We synthesized the electric field distributions using the measured transfer functions. Unlike the ideal free-space assumptions in simulations, these measured transfer functions encapsulate real-world physical factors, including mutual coupling between array elements, scattering from the feed support structure. The optimization algorithm was applied directly to this measured data to solve for the excitation coefficients required to generate plane waves with specific AoA. Fig. \ref{results} illustrates the synthesized normalized amplitude and phase distributions along the horizontal centerline of the QZ for four distinct incident azimuth angles: $0^\circ$, $3^\circ$, $7^\circ$, and $10^\circ$. To demonstrate the broadband capability of the system, results are overlaid for three representative frequencies: $2$ GHz, $3$ GHz, and $4$ GHz. 

The black dashed lines indicates the targeted $2$ m QZ width. For the normal incidence case ($0^\circ$), as shown in Fig. \ref{results} (a) and (b), the synthesized field exhibits a flat amplitude response within the QZ. The phase distribution remains essentially constant across the distinct frequencies, indicating a high-quality plane wave front perpendicular to the propagation direction. For the oblique incidence scenarios of $3^\circ$, $7^\circ$ and $10^\circ$, the amplitude distributions shown in Fig. \ref{results} (c), (e) and (g) maintain varying degrees of flatness within the QZ. Correspondingly, the phase distributions shown in Fig. \ref{results} (d), (f) and (g) display clear linear gradients. The slope of these gradients increases with the incident angle, aligning precisely with the theoretical phase progression required for beam steering. A quantitative summary of the measured field quality is presented in Table \ref{tab:measured_results}, which lists the peak-to-peak amplitude ripple and phase error within the QZ. {We can observe that the amplitude ripple and phase error within the QZ exhibits a gradual increase with the increase of incident angles. When emulating the target with $10^\circ$ incident angle at 2 GHz, the peak-to-peak phase error is $11.2^\circ$. However, in OTA testing systems (e.g., guided by 3GPP TR 38.810 standard practices \cite{3gpp.38.810.v16.7.0}), the typical acceptance criteria require the QZ amplitude ripples control under $\pm 0.5$ dB and phase ripple under $\pm 5$ degree. Even for commercial and mature OTA setup, it is difficult to maintain such QZ quality \cite{Keysight_F9650A_online}. Therefore, while an $11.2^\circ$ phase error at the extreme $10^\circ$ incident angle is on the higher end, it still falls within the acceptable operational envelope for general system-level evaluation. Moreover, a peak-to-peak amplitude and phase ripple of 1 dB and 10 degrees will not severely affect the high resolution DoA algorithms, because the peak-to-peak error will be averaged out by the array elements} The overall QZ quality remains robust across the investigated $2$--$4$ GHz frequency band. The consistently low ripple and error values at distinct frequencies effectively validate the proposed system's capability to support wideband testing.

\begin{table}[!t]
	\centering
	\caption{Summary of Field Quality in the Quiet Zone}
	\label{tab:measured_results}
	\renewcommand{\arraystretch}{1.3} 
	
	\begin{tabularx}{\columnwidth}{Y Y Y Y}
		\toprule
		\textbf{Frequency } & 
		\textbf{Incident Angle} & 
		\textbf{Amplitude peak-to-peak} & 
		\textbf{Phase peak-to-peak} \\
		\midrule
		
		\multirow{4}{*}{2 GHz} & 0$^\circ$ & 0.1 dB & 1.2$^\circ$ \\
		& 3$^\circ$ & 0.5 dB & 3.0$^\circ$ \\
		& 7$^\circ$ & 0.9 dB & 5.1$^\circ$ \\
		& 10$^\circ$ & 0.9 dB & 11.2.$^\circ$ \\
		\midrule
		\multirow{4}{*}{3 GHz} & 0$^\circ$ & 0.4 dB & 4.3$^\circ$ \\
		& 3$^\circ$ & 0.7 dB & 5.3$^\circ$ \\
		& 7$^\circ$ & 0.9 dB& 8.0$^\circ$ \\
		& 10$^\circ$ & 1.4 dB & 6.1$^\circ$ \\
		\midrule
		\multirow{4}{*}{4 GHz} & 0$^\circ$ & 0.2 dB & 2.2$^\circ$ \\
		& 3$^\circ$ & 0.5 dB & 3.6$^\circ$ \\
		& 7$^\circ$ & 0.8 dB & 5.6$^\circ$ \\
		& 10$^\circ$ & 1.5 dB & 6.6$^\circ$ \\
		\bottomrule
	\end{tabularx}
\end{table}

Furthermore, it is important to note that due to the inherent geometric symmetry of the SPCCR and the linear feed array with respect to the central axis, the performance characteristics observed for positive incident angles ($0^\circ$ to $10^\circ$) are representative of the negative angular domain ($0^\circ$ to $-10^\circ$). Consequently, the experimental validation presented here effectively demonstrates a total angular coverage of approximately $20^\circ$ (spanning $\pm 10^\circ$), verifying the system's capability to emulate closely-spaced spatial targets with high precision.

\section{Conclusion}
This paper proposes a novel OTA testing system based on a linear array feed SPCCR for the high-fidelity emulation of sensing targets, specifically addressing the challenge of resolving closely-spaced targets in the spatial domain. By optimizing the complex excitation coefficients of the linear array feed, the proposed system overcomes the limitations of traditional test system, which typically lack physical spatial resolution, and enables the synthesis of plane waves with precise AoA within a constrained compact range environment. Extensive FEKO simulations validated the system's capability to emulate multiple target under various incident angles, while preserving high quality QZ with amplitude ripples below $1.5$ dB and phase errors under $10^\circ$. An end-to-end simulation incorporating an ISAC BS model verifies the system's capability to accurately generate multiple spatial targets. To verify the practical feasibility, experimental validations were conducted using measured transfer functions across a frequency band of $2–4$ GHz.

In future research, we intend to extend the experimental validation to mmWave bands to fully exploit the ultra-wideband capability of the reflector. Additionally, end-to-end testing with commercial RTS and practical ISAC BSs will be conducted to further evaluate the system's efficacy in realistic sensing and communication scenarios.

\bibliographystyle{IEEEtranchange} 
\bibliography{references}

\newpage
\vfill

\end{document}